\documentclass[aps,pra,twocolumn,showpacs,groupedaddress,superscriptaddress,amsmath,amssymb]{revtex4}
\usepackage{graphicx}% Include figure files
\usepackage{natbib}
\begin{document}

\newcommand{\Fig}{Fig.~}
\newcommand{\Figs}{Figs.~}
\newcommand{\Figfirst}{Figure~}
\newcommand{\Rb}{${}^{87}$Rb}
\newcommand{\Rbe}{${}^{85}$Rb}
\newcommand{\fxm}[1]{{\bf !---!FIXME #1 !---!}}
\newcommand{\ds}{}
\newcommand{\etal}{\textit{et al.~}}
\newcommand{\ChapSecRef}[2]{Chapter #1 section #2}
\newcommand{\bra}[1]{ \langle #1 |}
\newcommand{\ket}[1]{ | #1 \rangle }
\newcommand{\hideit}[1]{ }
%%%%%%%%
% Acronyms
\newcommand{\EIT}{
	electro-magnetically induced transparency (EIT)
	\renewcommand{\EIT}{ EIT }
}

\newcommand{\EIA}{
	electro-magnetically induced absorption (EIA)
	\renewcommand{\EIA}{ EIA }
}

\newcommand{\BGIA}{
	buffer gas induced absorption (BGIA)
	\renewcommand{\BGIA}{ BGIA }
}

\newcommand\PN{
	phase noise (PN)
	\renewcommand\PN{ PN }
}

\newcommand\AOM{
	acousto-optic modulator (AOM)
	\renewcommand\AOM{ AOM }
}

\newcommand\EOM{
	electro-optic modulator (EOM)
	\renewcommand\EOM{ EOM }
}

\newcommand\FWHM{
	full width half maximum (FWHM)
	\renewcommand\FWHM{ FWHM }
}

%{{{ authors
\author{Eugeniy E.\ Mikhailov}
    \email{evmik@leona.physics.tamu.edu}
\affiliation{
	Department of Physics and Institute for Quantum Studies,
	Texas A\&M University,
	College Station, TX 77843
}
\affiliation{
	Center for Space Research,
	Massachusetts Institute of Technology,
	Cambridge, MA 02139
}
%   \homepage{http://leona.physics.tamu.edu/~evmik/}
\author{Vladimir A.\ Sautenkov}
\affiliation{
	Department of Physics and Institute for Quantum Studies,
	Texas A\&M University,
	College Station, TX 77843
}
\affiliation{
	Lebedev Institute of Physics, 119991 Moscow, Russia
}
\author{Irina Novikova}
\affiliation{
	Department of Physics and Institute for Quantum Studies,
	Texas A\&M University,
	College Station, TX 77843
}
\affiliation{
	Harvard-Smithsonian Center for Astrophysics,
	Cambridge, MA 02138 
}
\author{George R.\ Welch}
\affiliation{
	Department of Physics and Institute for Quantum Studies,
	Texas A\&M University,
	College Station, TX 77843
}
%\thanks is used only if any from above is not applied
%\thanks{}
%}}}

\title{ %{{{
	Large negative and positive delay of optical pulses
	in coherently prepared dense Rb vapor with buffer gas.
} %}}}

\begin{abstract} %{{{
We experimentally study the group time delay for a light
pulse propagating through hot \Rb{} vapor in the presence
of a strong coupling field in a $\Lambda$ configuration.
We demonstrate that the ultra-slow pulse propagation is
transformed into superluminal propagation as the one-photon
detuning of the light increases due to the change in the
transmission resonance lineshape.  Negative group velocity
as low as $-c/10^6=-80~\mathrm{m/s}$ is recorded. We also
find that the advance time in the regime of the superluminal
propagation grows linearly with increasing laser field power.
\end{abstract}
%}}}

\date{\today}
\maketitle
%%%%%%%%%%%%%%%%%%%%%%%%%%%%%%%%%%%%%%%%%%%%%%%%%%%%%%%%%%%%%%%%%%%%%%%%%%%

\section{Introduction} %{{{

	Coherent population trapping (CPT) occurs when
three-level atoms interact with two coherent electromagnetic
fields in a $\Lambda$ configuration.  In this case, atoms
are optically pumped into the non-interacting superposition
of the two ground states (into the ``dark'' state), and the
linear absorption of the optical fields is greatly reduced.
In addition to the absorption cancellation, such atomic media
exhibit steep nonlinear dispersion.  This steep dispersion can
result in significant reduction of the group velocity $v_g$
of a pulse propagating in a coherent medium, even though the
refractive index $n$ is still very close to unity:
\begin{equation}
v_g=\frac{c}{n+{\omega}\frac{{\partial}n}{{\partial}{\omega}}}~.
\end{equation}
Group velocity as low as a few meters per second has been
demonstrated experimentally in alkali atoms in Bose-Einstein
condensates~\cite{hau99}, thermal atomic vapor~\cite{kash99,budker99}, 
ruby crystal at room temperature~\cite{Bigelow03prl} 
and in Pr doped $\mathrm{Y_2SiO_5}$
crystals~\cite{TurukhinSSMHH02}.  See Ref.~\cite{MatskoKRWZS01}
for a comprehensive review of slow light-related research.

	The amplitude and width of the transmission and
dispersion resonances, and therefore the nonlinear properties
of the medium, are largely determined by the time atoms stay in
the dark state.  In thermal alkali vapors this time is limited
by the interaction time of the atoms with the laser fields,
which in turn is usually limited by the average time-of-flight
of an atom through the laser beam.  To prolong the interaction
time, an inert buffer gas is often added to the atomic vapor to
slow the diffusion of the coherently prepared atoms through the
laser beam~\cite{brandt'97, wynands'98, helm'01}.  However, it
has been recently demonstrated that significant modification of
the dark resonance is possible in atomic vapor in the presence
of buffer gas~\cite{mikhailov'03prep, mikhailov03praprep}.
Namely, the initially symmetric narrow transmission resonance
is transformed into an asymmetric dispersion-like lineshape and
then into a (potentially narrower) nearly symmetric absorption
resonance as the one-photon detuning of the laser fields
from the excited state is increased.  Therefore a far-detuned
$\Lambda$ system may exhibit superluminal propagation of the
probe pulse due to the steep anomalous dispersion associated
with a narrow absorption line.

	Although the superluminal propagation of a light pulse
near resonant absorption or gain lines has been known and
discussed for more than a century~\cite{chiao_book}, interest
in this effect has been revived recently by the observation
of ultra-low negative group velocity~\cite{wang2000nature,
wang2001pra, kuzmich2001prl, godone02pra, akulshin03pra, 
akulshin03joptb, Bigelow03sci, stenner2003nat, kim2003pra}.  
Here we present an extensive study
of the time delay of probe pulses propagating in a ${\Lambda}$
scheme under the conditions of CPT.  We trace the modification
of positive delay for zero one-photon detuning to negative
delay for large one-photon detuning.  Also we present the
dependence of the delay time on total laser power.
%
%}}}

\section{Experimental setup} %{{{

\begin{figure}
\includegraphics[angle=00,
	%height=0.85\textheight,
	width=1.00\columnwidth
	]
	{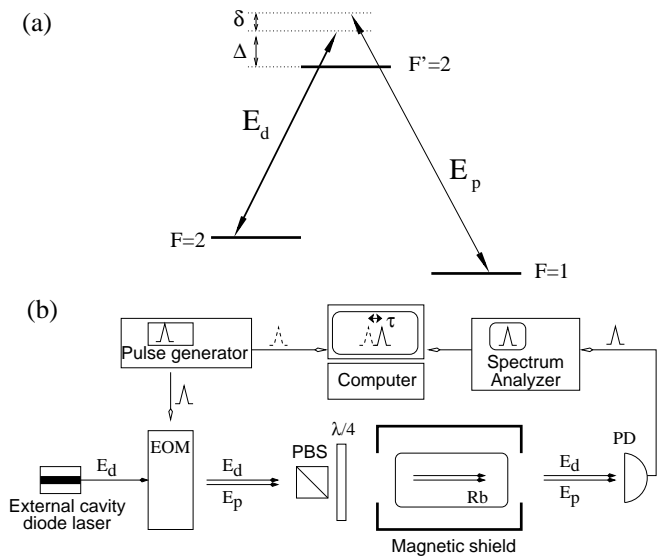}
\caption{
	(a) \Rb{} level diagram;
	(b) Schematic of the experimental apparatus.
	\label{setup_and_levels}
}
\end{figure}

	A schematic of the experimental setup is shown in
\Fig\ref{setup_and_levels}b.  An extended cavity diode laser
is tuned to $5^2S_{1/2} F=2 \to 5^2P_{1/2} F'=2$ transition of
\Rb{} ($\lambda=795$~nm).  The laser output is phase modulated
using an electro-optic modulator (EOM) at a frequency close
to the \Rb{} ground-state splitting (6.835~GHz), so that one
of the modulation sidebands is resonant with $5^2S_{1/2} F=1
\to 5^2P_{1/2} F=2$ transition, forming the $\Lambda$ system
shown in \Fig\ref{setup_and_levels}a.  The other sideband is
far-detuned from all \Rb{} transitions, and has no effect on the
phenomena described below.  Approximately $7\%$ of the total
laser power is transferred to the probe field.  After the EOM
the probe and drive fields pass through a single mode optical
fiber to ensure a Gaussian spatial intensity distribution,
and their polarization is made circular by a high-quality
polarizer followed by a quarter wave-plate.  The overall
light intensity is controlled by the rotation of an additional
polarizer placed after the fiber.  The diameter of the laser
beam at the entrance of \Rb{} cell is approximately $7$~mm.

	The experiment is conducted using a $2.5$~cm
long cylindrical glass cell filled with a mixture of
isotopically enhanced \Rb{} and 30 Torr of Ne buffer gas.
The cell is placed inside a 3 layer magnetic shield to
screen out the Earth's magnetic field, and maintained at
$68^o$~C, which results in an atomic density of \Rb{} vapor
$N=4.7{\times}10^{11}~\mathrm{cm}^{-3}$.  After the cell
the optical fields are mixed with an additional optical field
shifted by 60 MHz with respect to the original laser frequency,
and the amplitude of the beat signal between this field and
and the probe field is used to monitor the changes in the
transmitted probe field intensity.

%}}}

\section{Experimental results} %{{{

\begin{figure}
\includegraphics[
angle=00,
	%height=0.85\textheight,
	width=1.00\columnwidth
	]
	{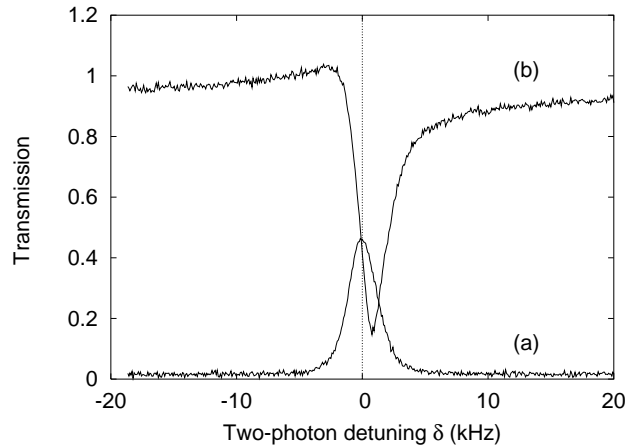}
\caption{
	The probe field transmission resonance for (a)
	zero one-photon detuning ${\Delta}=0$ and (b)
	for ${\Delta}=1.45$~GHz.  The output signals
	are shown on the same scale and normalized to the
	\textit{non-absorbed} transmission.  The total power
	of the laser beam is 400~${\mu}$W.
\label{resonance_shape} }
\end{figure}

	To observe the modification of the dark resonance
lineshape we change the one-photon detuning $\Delta$ by
changing the laser frequency, and then record the probe
field transmission as a function of two-photon detuning by
scanning the microwave frequency driving the EOM.  In the
case of a nearly resonant $\Lambda$ system ($\Delta=0$),
a narrow transmission resonance is observed due to coherent
population trapping.  An example of such a dark resonance
is shown in \Fig~\ref{resonance_shape}a.  As the one-photon
detuning increases, the shape of the resonance changes, and
for large detuning it becomes almost purely absorptive (see
\Fig~\ref{resonance_shape}b).  Note that the widths of both
resonances are similar ($\approx 2.5$~kHz) and are determined
by the decay rate of the ground-state atomic coherence.

\begin{figure}
\center
\includegraphics[angle=00,
%height=1.00\textheight,
width=1.00\columnwidth
]
	{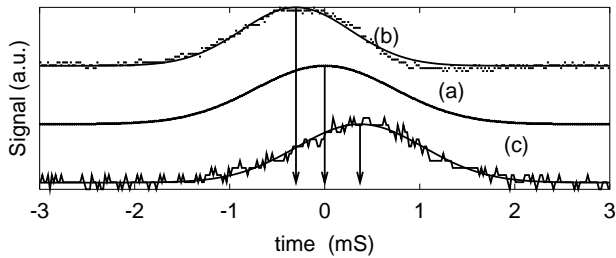}
	\caption{
		Examples of transmitted pulses:
		(a) reference pulse (no atoms);
		(b) advanced pulse recorded at one-photon
		detuning ${\Delta}=1.45$~GHz (total laser
		power was 700~${\mu}$W);
		(c) retarded pulse recorded at ${\Delta}=0$
		(total laser power 145~${\mu}$W).  Solid
		lines represents the Gaussian lineshape.  The
		amplitudes of all pulses are normalized for
		easier comparison.
\label{pulse_delay_example}
}
\end{figure}

	The corresponding changes in the atomic dispersion
from normal ($\partial n/\partial \omega >0$) to anomalous
($\partial n/\partial \omega <0$) produces a modification
of the pulse delay from positive (slow light) to negative
(superluminal regime, or fast light).  Examples of such pulses
are shown in \Fig\ref{pulse_delay_example}.  As expected the
transmitted pulse for the resonant probe field is delayed with
respect to the reference for ${\tau}=370~{\mu}$s, whereas for
the far-detuned $\Lambda$ system the maximum of the output
pulse leaves the cell noticeably earlier than the input.
For the laser detuning of ${\Delta}=1.45$~GHz the advance
time is 300~${\mu}$s, which corresponds to a group index
$n_g\simeq{-4}{\times}10^6$.

	It is easy to see that only minimal reshaping of both
retarded and advanced probe pulses is observed.  In either case,
the output pulses are still very close to the original Gaussian
waveform, although their widths are slightly reduced.  For a
1~ms duration input probe pulse the duration of the transmitted
pulse is $0.94$~ms in the case of slow light propagation and
$0.81$~ms in the case of superluminal propagation.

\begin{figure}
\includegraphics[angle=00,
	%height=0.85\textheight,
	width=1.00\columnwidth
	]
	{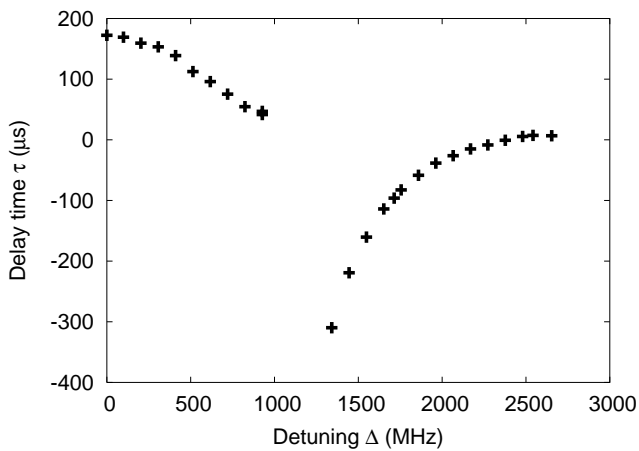}
\caption{
	Dependence of the positive and negative pulse delays
	on one photon laser detuning.  Total power of the
	laser beam is 400~${\mu}$W.
   \label{delay_vs_detuning}
}
\end{figure}

	It is interesting to track the variation of the group
velocity as the shape of the dark resonance changes from
transmission to absorption with increasing laser detuning.
This dependence of the delay time on the laser frequency is
shown in \Fig\ref{delay_vs_detuning}.  This dependence is
measured in the following way:  for each one photon detuning
we adjust the the two-photon tuning of the probe field to the
transmission peak (maximum or minimum) by adjusting the EOM
modulation frequency.  Then we propagate a Gaussian temporal
probe field pulse through the medium and measure its relative
delay with respect to the reference pulse.  Close to atomic
resonance ($\Delta <1$~GHz) the EIT transmission peak is
observed, although it becomes asymmetric as the laser detuning
increases.  The gap in measurements between 1~GHz and 1.4~GHz
corresponds to the range where the dark resonance lineshape
is too dispersion-like to find a meaningful peak.  Although
theoretically both slow and superluminal group velocities
might be measured in this case for two slightly different
two-photon detunings, in practice it is rather difficult
to perform such measurements.  In this regime the measured
group velocity is extremely sensitive to any changes in the
one-photon detuning, and therefore the instability of the laser
frequency creates a huge variation of the measured pulse delays.
Once the absorption resonance begins to dominate (for $\Delta >
1.4$~GHz), we follow the minimum of the probe transmission by
again adjusting the two-photon detuning appropriately.

	We have also studied the dependence of the group
velocity on the drive field power.  We measure the pulse delay
both for near-resonant (EIT) and far-detuned fields (enhanced
absorption resonance) as the laser intensity is changed.
The results are shown in \Fig\ref{resonance_shape}.  One can
see that in the case of slow pulse propagation the group
delay is inversely proportional to the laser intensity. This
dependence is well described by the power broadening of the
dark resonance~\cite{kash99}:
\begin{equation}
\label{tau_slow}
\tau \simeq \frac{3}{8\pi}N\lambda^2L\frac{\gamma_r}{| \Omega| ^ 2}~.
\end{equation}
This assumes the usual EIT conditions, namely
$|\Omega|^2 \gg \gamma_0\gamma$ and $|\Omega|^2 \gg
\sqrt{\gamma_0/\gamma}\,W_D$, where $\Omega$ is the drive field
Rabi frequency, $\gamma_r$ and $\gamma$ are the radiative and
total decay rates of the excited states, $W_D$ is the width
of the Doppler-broadened absorption line, and $\gamma_0$
is the dark state decoherence rate.

\begin{figure} \includegraphics[angle=00,
%height=0.85\textheight,
width=1.00\columnwidth 	]
{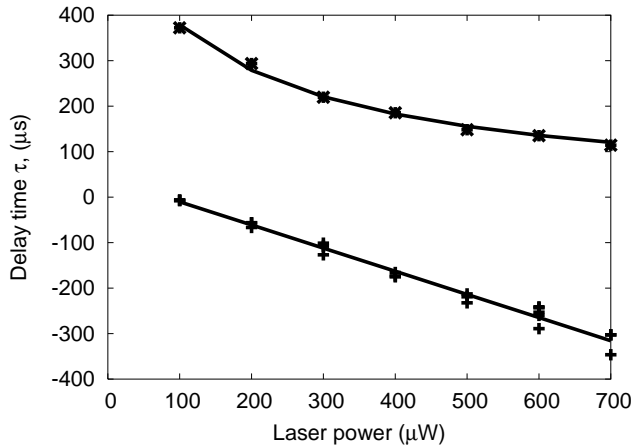}
\caption{
	The dependence of the positive and negative pulse delays
	on the laser power.  The two graphs are for $\Delta = 0$
	(x) and $\Delta = 1.44$~GHz (+).  Solid lines represent
	fits to the experimental data for the two cases:
	if the lasers are tuned to the atomic resonances,
	$\tau \propto 1/|\Omega|^2$, and for the far-detuned
	$\Lambda$ system $\tau\propto -|\Omega|^2$.
\label{delay_vs_pover}
}
\end{figure}

\begin{figure}
\includegraphics[angle=00,
	%height=0.85\textheight,
	width=1.00\columnwidth
]
{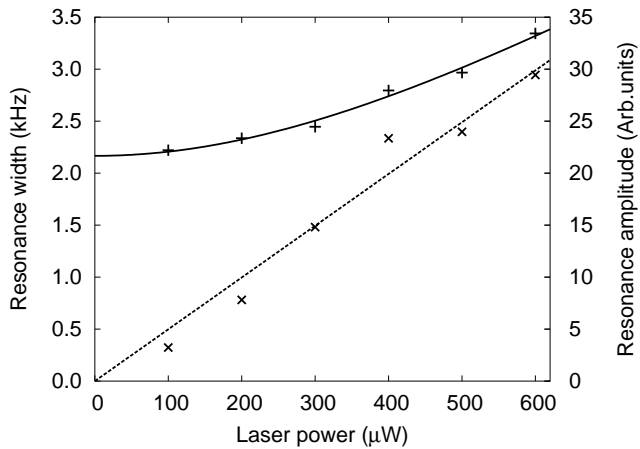}
\caption{
	The width (+) and the amplitude (x) of the absorption
	resonance as a function of the laser power.  The data
	are taken for one photon detuning $\Delta = 1.44$~GHz.
\label{widthandamp_vs_power}
}
\end{figure}

	Very different behavior is observed for the superluminal
pulse propagation.  Namely, the advance time linearly increases
with the laser power, which means that the steepness of
the anomalous dispersion increases as well.  This behavior may
be explained by the fact that for the far-detuned $\Lambda$
system the width of the absorption resonance is much less
affected by power broadening.  The asymptotic behavior
may be deduced from the general expression for the dark
resonance width~\cite{mikhailov03praprep}
\begin{equation}
\label{width}
\gamma_{\scriptscriptstyle\mathrm{EIT}} \simeq
 \gamma_0+\frac{\gamma^2|\Omega|^4}{2\gamma_0\Delta^4}
\end{equation}
The experimental measurements of the resonance width vs.\
laser power are shown in Fig.~\ref{widthandamp_vs_power} and
are in good agreement with this expression.  At the same time,
the amplitude of the resonance is directly proportional to the
laser power.  The combination of these two effects provides
the linear growth of the atomic dispersion and the probe pulse
advance time.

	We can check this conclusion in the following way.
The nonlinear dispersion of the atomic medium may be calculated
analytically using the well-known density matrix equations
for a three-level $\Lambda$ system, if the motion of the
atoms is neglected.  Following the calculations presented
in~\cite{mikhailov03praprep} we find the time delay is given by:
\begin{eqnarray}
\tau &=& \frac{3}{8\pi}N\lambda^2L
\gamma_r\frac{\gamma^2+\Delta^2}{2\gamma_0\Delta^2+\gamma|\Omega|^2}
\times \nonumber \\
&&\frac{\gamma|\Omega|^4-2|\Omega|^2\displaystyle{\frac{\Delta^2}{\gamma^2+
\Delta ^2}}(\gamma_0\Delta^2+\gamma|\Omega|^2)}{
\gamma_0^2\Delta^2(\gamma^2+\Delta^2)+\gamma^2|\Omega|^4}
\end{eqnarray}
Although this expression is very cumbersome it reduces to
Eq.~(\ref{tau_slow}) at one-photon resonance ($\Delta=0$).
A simple asymptotic behavior of the time delay may be also found
for large detunings $\Delta^2 \gg \gamma/\gamma_0|\Omega|^2$
we find
\begin{equation}
\label{fast_tau}
\tau \simeq  -\frac{3}{8\pi}N\lambda^2L\frac{\gamma_r| \Omega| ^
	2}{\gamma_0^2\Delta^2}~.
\end{equation}
This linear increase in pulse advance time with laser power
is just what is seen in \Fig\ref{delay_vs_pover}.

%}}}

\section{Discussion}
%{{{

	As previously mentioned, both large positive and
negative group delay are the manifestation of steep atomic
dispersion, either normal or anomalous.  This means that the
accumulated phase of an electromagnetic field traversing such
a medium is very sensitive to even small variations in its
frequency.  Therefore, it may be measured with high precision
if phase-sensitive measurements of the output field are made.
The cancellation of the linear absorption is undoubtedly the
biggest advantage of the steep dispersion associated with
EIT, and has been widely exploited for precision magnetic
field measurements~\cite{scully92prl, fleischhauer94pra,
wynands'98, stahler01eurlett, budker'00, NovikovaW02} and atomic
standards~\cite{vanier98, hollberg'02, merimaa'03, Vanier03}.

	In this light, our experimental data suggest that the
narrow absorption resonances observed in buffered atomic cells
in a far-detuned $\Lambda$ system are also good candidates
for precision measurements.  We can estimate the phase
$\Delta\varphi$ acquired by the probe field in this regime as:
\begin{equation} \label{}
\Delta\varphi \approx \frac{\omega}{c}L
\frac{{\partial}n}{{\partial}{\omega}}\delta \approx
	\tau\gamma_{\scriptscriptstyle\mathrm{EIT}}
\end{equation}
where we assume that the detuning of the probe field from the
center of the absorption resonance $\delta$ is on the order of
the resonance width $\gamma_{\scriptscriptstyle\mathrm{EIT}}$\,.
Substituting the experimental value for the group advance
time $\tau\simeq 0.3$~ms and for the resonance width
$\gamma\simeq 2.5$~kHz, we find the maximum phase difference
to be about $5$~rad, which is on the same order of magnitude
as the large polarization rotation angle in nonlinear
Faraday rotation~\cite{novikova'01ol} and the large phase
shifts induced by gradient magnetic fields in stored light
experiments~\cite{phillips03}.

%}}}

\section{Conclusion} %{{{

	We have studied the propagation of a weak probe pulse
in an atomic cell filled with optically thick \Rb{} vapor and
Ne buffer gas for various one-photon detunings from the upper
atomic state of the $\Lambda$ system.  The pulse propagation
undergoes a transition from an ultra-slow to a superluminal
regime as the dark resonance lineshape changes from a narrow
transmission to a narrow absorption resonance.  In the latter
case, advance time as long as $300 \mu\mathrm{s}$ has been
observed, which is equivalent to a negative group velocity of
$v_g\simeq -80$~m/s.

	We have also studied the dependence of the group
delay on the laser power.  While the group velocity under
EIT conditions increases linearly, in agreement with the
theoretical predictions, we see a linear growth of negative
group delay with the laser power.  The main reason is that
for large one-photon detuning the contrast of the absorption
resonance increases with laser intensity, whereas the change
in the resonance width is negligible.

	The large anomalous dispersion associated with
the reported absorption resonances is associated with high
sensitivity of the phase of the probe field to the frequency
variations, which may be find application in precision
measurements.

%}}}

\begin{acknowledgments} %{{{

	The authors thank Yuri V.\ Rostovtsev, Andrey B.\
Matsko, A.\ Zhang and  M.\ O.\  Scully for useful discussions.
This  work was supported  by the the Office of Naval Research,
AFOSR, and the DARPA.

\end{acknowledgments}

%}}}

%%%%%%%%%%%%%%%%%%%%%%%%%%%%%%%%%%%%%%%%%%%%%%%%%%%%%%%%%%%%%%%%%%%%%%%%%%%

%%%%%%%%%%%%%%%%%%%%%%%%%%%%%%%%%%%%%%%%%%%%%%%%%%%%%%%%%%%%%%%%%%%%%%%%%%%


\begin{thebibliography}{32}
\expandafter\ifx\csname natexlab\endcsname\relax\def\natexlab#1{#1}\fi
\expandafter\ifx\csname bibnamefont\endcsname\relax
  \def\bibnamefont#1{#1}\fi
\expandafter\ifx\csname bibfnamefont\endcsname\relax
  \def\bibfnamefont#1{#1}\fi
\expandafter\ifx\csname citenamefont\endcsname\relax
  \def\citenamefont#1{#1}\fi
\expandafter\ifx\csname url\endcsname\relax
  \def\url#1{\texttt{#1}}\fi
\expandafter\ifx\csname urlprefix\endcsname\relax\def\urlprefix{URL }\fi
\providecommand{\bibinfo}[2]{#2}
\providecommand{\eprint}[2][]{\url{#2}}

\bibitem[{\citenamefont{Hau et~al.}(1999)\citenamefont{Hau, Harris, Dutton, and
  Behroozi}}]{hau99}
\bibinfo{author}{\bibfnamefont{L.~V.} \bibnamefont{Hau}},
  \bibinfo{author}{\bibfnamefont{S.~E.} \bibnamefont{Harris}},
  \bibinfo{author}{\bibfnamefont{Z.}~\bibnamefont{Dutton}}, \bibnamefont{and}
  \bibinfo{author}{\bibfnamefont{C.~H.} \bibnamefont{Behroozi}},
  \bibinfo{journal}{Nature} \textbf{\bibinfo{volume}{397}},
  \bibinfo{pages}{594} (\bibinfo{year}{1999}).

\bibitem[{\citenamefont{Kash et~al.}(1999)\citenamefont{Kash, Sautenkov,
  Zibrov, Hollberg, Welch, Lukin, Rostovtsev, Fry, and Scully}}]{kash99}
\bibinfo{author}{\bibfnamefont{M.~M.} \bibnamefont{Kash}},
  \bibinfo{author}{\bibfnamefont{V.~A.} \bibnamefont{Sautenkov}},
  \bibinfo{author}{\bibfnamefont{A.~S.} \bibnamefont{Zibrov}},
  \bibinfo{author}{\bibfnamefont{L.}~\bibnamefont{Hollberg}},
  \bibinfo{author}{\bibfnamefont{G.~R.} \bibnamefont{Welch}},
  \bibinfo{author}{\bibfnamefont{M.~D.} \bibnamefont{Lukin}},
  \bibinfo{author}{\bibfnamefont{Y.}~\bibnamefont{Rostovtsev}},
  \bibinfo{author}{\bibfnamefont{E.~S.} \bibnamefont{Fry}}, \bibnamefont{and}
  \bibinfo{author}{\bibfnamefont{M.~O.} \bibnamefont{Scully}},
  \bibinfo{journal}{Phys. Rev. Lett.} \textbf{\bibinfo{volume}{82}},
  \bibinfo{pages}{5229} (\bibinfo{year}{1999}).

\bibitem[{\citenamefont{Budker et~al.}(1999)\citenamefont{Budker, Kimball,
  Rochester, and Yashchuk}}]{budker99}
\bibinfo{author}{\bibfnamefont{D.}~\bibnamefont{Budker}},
  \bibinfo{author}{\bibfnamefont{D.~F.} \bibnamefont{Kimball}},
  \bibinfo{author}{\bibfnamefont{S.~M.} \bibnamefont{Rochester}},
  \bibnamefont{and} \bibinfo{author}{\bibfnamefont{V.~V.}
  \bibnamefont{Yashchuk}}, \bibinfo{journal}{Phys. Rev. Lett.}
  \textbf{\bibinfo{volume}{83}}, \bibinfo{pages}{1767} (\bibinfo{year}{1999}).

\bibitem[{\citenamefont{Bigelow
  et~al.}(2003{\natexlab{a}})\citenamefont{Bigelow, Lepeshkin, and
  Boyd}}]{Bigelow03prl}
\bibinfo{author}{\bibfnamefont{M.}~\bibnamefont{Bigelow}},
  \bibinfo{author}{\bibfnamefont{N.}~\bibnamefont{Lepeshkin}},
  \bibnamefont{and} \bibinfo{author}{\bibfnamefont{R.}~\bibnamefont{Boyd}},
  \bibinfo{journal}{Phys. Rev. Lett.} \textbf{\bibinfo{volume}{90}},
  \bibinfo{pages}{113903} (\bibinfo{year}{2003}{\natexlab{a}}).

\bibitem[{\citenamefont{Turukhin et~al.}(2002)\citenamefont{Turukhin,
  Sudarshanam, Shahriar, Musser, Ham, and Hemmer}}]{TurukhinSSMHH02}
\bibinfo{author}{\bibfnamefont{A.~V.} \bibnamefont{Turukhin}},
  \bibinfo{author}{\bibfnamefont{V.~S.} \bibnamefont{Sudarshanam}},
  \bibinfo{author}{\bibfnamefont{M.~S.} \bibnamefont{Shahriar}},
  \bibinfo{author}{\bibfnamefont{J.~A.} \bibnamefont{Musser}},
  \bibinfo{author}{\bibfnamefont{B.~S.} \bibnamefont{Ham}}, \bibnamefont{and}
  \bibinfo{author}{\bibfnamefont{P.~R.} \bibnamefont{Hemmer}},
  \bibinfo{journal}{Phys. Rev. Lett.} \textbf{\bibinfo{volume}{88}},
  \bibinfo{pages}{023602} (\bibinfo{year}{2002}).

\bibitem[{\citenamefont{Matsko et~al.}(2001)\citenamefont{Matsko,
  Kocharovskaya, Rostovtsev, Welch, Zibrov, and Scully}}]{MatskoKRWZS01}
\bibinfo{author}{\bibfnamefont{A.~B.} \bibnamefont{Matsko}},
  \bibinfo{author}{\bibfnamefont{O.}~\bibnamefont{Kocharovskaya}},
  \bibinfo{author}{\bibfnamefont{Y.}~\bibnamefont{Rostovtsev}},
  \bibinfo{author}{\bibfnamefont{G.~R.} \bibnamefont{Welch}},
  \bibinfo{author}{\bibfnamefont{A.~S.} \bibnamefont{Zibrov}},
  \bibnamefont{and} \bibinfo{author}{\bibfnamefont{M.~O.}
  \bibnamefont{Scully}}, \bibinfo{journal}{Advan Atom Mol Opt Phys}
  \textbf{\bibinfo{volume}{46}}, \bibinfo{pages}{191} (\bibinfo{year}{2001}).

\bibitem[{\citenamefont{Brandt et~al.}(1997)\citenamefont{Brandt, Nagel,
  Wynands, and Meschede}}]{brandt'97}
\bibinfo{author}{\bibfnamefont{S.}~\bibnamefont{Brandt}},
  \bibinfo{author}{\bibfnamefont{A.}~\bibnamefont{Nagel}},
  \bibinfo{author}{\bibfnamefont{R.}~\bibnamefont{Wynands}}, \bibnamefont{and}
  \bibinfo{author}{\bibfnamefont{D.}~\bibnamefont{Meschede}},
  \bibinfo{journal}{Phys. Rev. A} \textbf{\bibinfo{volume}{56}},
  \bibinfo{pages}{R1063} (\bibinfo{year}{1997}).

\bibitem[{\citenamefont{Wynands and Nagel}(1998)}]{wynands'98}
\bibinfo{author}{\bibfnamefont{R.}~\bibnamefont{Wynands}} \bibnamefont{and}
  \bibinfo{author}{\bibfnamefont{A.}~\bibnamefont{Nagel}},
  \bibinfo{journal}{Appl.\ Phys.\ B} \textbf{\bibinfo{volume}{68}},
  \bibinfo{pages}{1 } (\bibinfo{year}{1998}).

\bibitem[{\citenamefont{Erhard and Helm}(2001)}]{helm'01}
\bibinfo{author}{\bibfnamefont{M.}~\bibnamefont{Erhard}} \bibnamefont{and}
  \bibinfo{author}{\bibfnamefont{H.}~\bibnamefont{Helm}},
  \bibinfo{journal}{Phys. Rev. A} \textbf{\bibinfo{volume}{63}},
  \bibinfo{pages}{043813} (\bibinfo{year}{2001}).

\bibitem[{\citenamefont{Mikhailov
  et~al.}(2003{\natexlab{a}})\citenamefont{Mikhailov, Sautenkov, Rostovtsev,
  and Welch}}]{mikhailov'03prep}
\bibinfo{author}{\bibfnamefont{E.~E.} \bibnamefont{Mikhailov}},
  \bibinfo{author}{\bibfnamefont{V.~A.} \bibnamefont{Sautenkov}},
  \bibinfo{author}{\bibfnamefont{Y.~V.} \bibnamefont{Rostovtsev}},
  \bibnamefont{and} \bibinfo{author}{\bibfnamefont{G.~R.} \bibnamefont{Welch}},
  \bibinfo{journal}{LANL e-Print archive}
  (\bibinfo{year}{2003}{\natexlab{a}}),
  \urlprefix\url{http://arxiv.org/abs/quant-ph/0309151}.

\bibitem[{\citenamefont{Mikhailov
  et~al.}(2003{\natexlab{b}})\citenamefont{Mikhailov, Novikova, Rostovtsev, and
  Welch}}]{mikhailov03praprep}
\bibinfo{author}{\bibfnamefont{E.~E.} \bibnamefont{Mikhailov}},
  \bibinfo{author}{\bibfnamefont{I.}~\bibnamefont{Novikova}},
  \bibinfo{author}{\bibfnamefont{Y.~V.} \bibnamefont{Rostovtsev}},
  \bibnamefont{and} \bibinfo{author}{\bibfnamefont{G.~R.} \bibnamefont{Welch}},
  \bibinfo{journal}{LANL e-Print archive}
  (\bibinfo{year}{2003}{\natexlab{b}}),
  \urlprefix\url{http://arxiv.org/abs/quant-ph/0309171}.

\bibitem[{\citenamefont{Chiao}(1996)}]{chiao_book}
\bibinfo{author}{\bibfnamefont{R.~Y.} \bibnamefont{Chiao}},
  \emph{\bibinfo{title}{Amazing Light}} (\bibinfo{publisher}{Springer, New
  York}, \bibinfo{year}{1996}).

\bibitem[{\citenamefont{Wang et~al.}(2000)\citenamefont{Wang, Kuzmich, and
  Dogariu}}]{wang2000nature}
\bibinfo{author}{\bibfnamefont{L.~J.} \bibnamefont{Wang}},
  \bibinfo{author}{\bibfnamefont{A.}~\bibnamefont{Kuzmich}}, \bibnamefont{and}
  \bibinfo{author}{\bibfnamefont{A.}~\bibnamefont{Dogariu}},
  \bibinfo{journal}{Nature} \textbf{\bibinfo{volume}{406}},
  \bibinfo{pages}{277} (\bibinfo{year}{2000}).

\bibitem[{\citenamefont{Dogariu et~al.}(2001)\citenamefont{Dogariu, Kuzmich,
  and Wang}}]{wang2001pra}
\bibinfo{author}{\bibfnamefont{A.}~\bibnamefont{Dogariu}},
  \bibinfo{author}{\bibfnamefont{A.}~\bibnamefont{Kuzmich}}, \bibnamefont{and}
  \bibinfo{author}{\bibfnamefont{L.~J.} \bibnamefont{Wang}},
  \bibinfo{journal}{Phys. Rev. A} \textbf{\bibinfo{volume}{63}},
  \bibinfo{pages}{053806} (\bibinfo{year}{2001}).

\bibitem[{\citenamefont{Kuzmich et~al.}(2001)\citenamefont{Kuzmich, Dogariu,
  Wang, Milonni, and Chiao}}]{kuzmich2001prl}
\bibinfo{author}{\bibfnamefont{A.}~\bibnamefont{Kuzmich}},
  \bibinfo{author}{\bibfnamefont{A.}~\bibnamefont{Dogariu}},
  \bibinfo{author}{\bibfnamefont{L.~J.} \bibnamefont{Wang}},
  \bibinfo{author}{\bibfnamefont{P.~W.} \bibnamefont{Milonni}},
  \bibnamefont{and} \bibinfo{author}{\bibfnamefont{R.~Y.} \bibnamefont{Chiao}},
  \bibinfo{journal}{Phys. Rev. Lett.} \textbf{\bibinfo{volume}{86}},
  \bibinfo{pages}{3925} (\bibinfo{year}{2001}).

\bibitem[{\citenamefont{Godone et~al.}(2002)\citenamefont{Godone, Levi, and
  Micalizio}}]{godone02pra}
\bibinfo{author}{\bibfnamefont{A.}~\bibnamefont{Godone}},
  \bibinfo{author}{\bibfnamefont{F.}~\bibnamefont{Levi}}, \bibnamefont{and}
  \bibinfo{author}{\bibfnamefont{S.}~\bibnamefont{Micalizio}},
  \bibinfo{journal}{Phys. Rev. A} \textbf{\bibinfo{volume}{66}},
  \bibinfo{pages}{043804} (\bibinfo{year}{2002}).

\bibitem[{\citenamefont{Akulshin
  et~al.}(2003{\natexlab{a}})\citenamefont{Akulshin, Cimmino, Sidorov,
  Hannaford, and Opat}}]{akulshin03pra}
\bibinfo{author}{\bibfnamefont{A.~M.} \bibnamefont{Akulshin}},
  \bibinfo{author}{\bibfnamefont{A.}~\bibnamefont{Cimmino}},
  \bibinfo{author}{\bibfnamefont{A.~I.} \bibnamefont{Sidorov}},
  \bibinfo{author}{\bibfnamefont{P.}~\bibnamefont{Hannaford}},
  \bibnamefont{and} \bibinfo{author}{\bibfnamefont{G.~I.} \bibnamefont{Opat}},
  \bibinfo{journal}{Phys. Rev. A} \textbf{\bibinfo{volume}{67}},
  \bibinfo{pages}{011801(R)} (\bibinfo{year}{2003}{\natexlab{a}}).

\bibitem[{\citenamefont{Akulshin
  et~al.}(2003{\natexlab{b}})\citenamefont{Akulshin, Cimmino, Sidorov, McLean,
  and Hannaford}}]{akulshin03joptb}
\bibinfo{author}{\bibfnamefont{A.~M.} \bibnamefont{Akulshin}},
  \bibinfo{author}{\bibfnamefont{A.}~\bibnamefont{Cimmino}},
  \bibinfo{author}{\bibfnamefont{A.~I.} \bibnamefont{Sidorov}},
  \bibinfo{author}{\bibfnamefont{R.}~\bibnamefont{McLean}}, \bibnamefont{and}
  \bibinfo{author}{\bibfnamefont{P.}~\bibnamefont{Hannaford}},
  \bibinfo{journal}{J. Opt. B: Quantum Semiclass. Opt.}
  \textbf{\bibinfo{volume}{5}}, \bibinfo{pages}{S479}
  (\bibinfo{year}{2003}{\natexlab{b}}).

\bibitem[{\citenamefont{Bigelow
  et~al.}(2003{\natexlab{b}})\citenamefont{Bigelow, Lepeshkin, and
  Boyd}}]{Bigelow03sci}
\bibinfo{author}{\bibfnamefont{M.}~\bibnamefont{Bigelow}},
  \bibinfo{author}{\bibfnamefont{N.}~\bibnamefont{Lepeshkin}},
  \bibnamefont{and} \bibinfo{author}{\bibfnamefont{R.}~\bibnamefont{Boyd}},
  \bibinfo{journal}{Science} \textbf{\bibinfo{volume}{301}},
  \bibinfo{pages}{200} (\bibinfo{year}{2003}{\natexlab{b}}).

\bibitem[{\citenamefont{Stenner et~al.}(2003)\citenamefont{Stenner, Gauthier,
  and Neifeld}}]{stenner2003nat}
\bibinfo{author}{\bibfnamefont{M.}~\bibnamefont{Stenner}},
  \bibinfo{author}{\bibfnamefont{D.}~\bibnamefont{Gauthier}}, \bibnamefont{and}
  \bibinfo{author}{\bibfnamefont{M.}~\bibnamefont{Neifeld}},
  \bibinfo{journal}{Nature} \textbf{\bibinfo{volume}{425}},
  \bibinfo{pages}{695} (\bibinfo{year}{2003}).

\bibitem[{\citenamefont{Kim et~al.}(2003)\citenamefont{Kim, Moon, Lee, Kim, and
  Kim}}]{kim2003pra}
\bibinfo{author}{\bibfnamefont{K.}~\bibnamefont{Kim}},
  \bibinfo{author}{\bibfnamefont{H.}~\bibnamefont{Moon}},
  \bibinfo{author}{\bibfnamefont{C.}~\bibnamefont{Lee}},
  \bibinfo{author}{\bibfnamefont{S.}~\bibnamefont{Kim}}, \bibnamefont{and}
  \bibinfo{author}{\bibfnamefont{J.}~\bibnamefont{Kim}},
  \bibinfo{journal}{Phys. Rev. A} \textbf{\bibinfo{volume}{68}},
  \bibinfo{pages}{013810} (\bibinfo{year}{2003}).

\bibitem[{\citenamefont{Scully and Fleischhauer}(1992)}]{scully92prl}
\bibinfo{author}{\bibfnamefont{M.~O.} \bibnamefont{Scully}} \bibnamefont{and}
  \bibinfo{author}{\bibfnamefont{M.}~\bibnamefont{Fleischhauer}},
  \bibinfo{journal}{Phys. Rev. Lett.} \textbf{\bibinfo{volume}{69}},
  \bibinfo{pages}{1360} (\bibinfo{year}{1992}).

\bibitem[{\citenamefont{Fleischhauer and Scully}(1994)}]{fleischhauer94pra}
\bibinfo{author}{\bibfnamefont{M.}~\bibnamefont{Fleischhauer}}
  \bibnamefont{and} \bibinfo{author}{\bibfnamefont{M.~O.}
  \bibnamefont{Scully}}, \bibinfo{journal}{Phys. Rev. A}
  \textbf{\bibinfo{volume}{49}}, \bibinfo{pages}{1973} (\bibinfo{year}{1994}).

\bibitem[{\citenamefont{Stahler et~al.}(2001)\citenamefont{Stahler, Knappe,
  Affolderbach, Kemp, and Wynands}}]{stahler01eurlett}
\bibinfo{author}{\bibfnamefont{M.}~\bibnamefont{Stahler}},
  \bibinfo{author}{\bibfnamefont{S.}~\bibnamefont{Knappe}},
  \bibinfo{author}{\bibfnamefont{C.}~\bibnamefont{Affolderbach}},
  \bibinfo{author}{\bibfnamefont{W.}~\bibnamefont{Kemp}}, \bibnamefont{and}
  \bibinfo{author}{\bibfnamefont{R.}~\bibnamefont{Wynands}},
  \bibinfo{journal}{Europhysics Lett.} \textbf{\bibinfo{volume}{54}},
  \bibinfo{pages}{323} (\bibinfo{year}{2001}).

\bibitem[{\citenamefont{Budker et~al.}(2000)\citenamefont{Budker, Kimball,
  Rochester, Yashchuk, and Zolotorev}}]{budker'00}
\bibinfo{author}{\bibfnamefont{D.}~\bibnamefont{Budker}},
  \bibinfo{author}{\bibfnamefont{D.~F.} \bibnamefont{Kimball}},
  \bibinfo{author}{\bibfnamefont{S.~M.} \bibnamefont{Rochester}},
  \bibinfo{author}{\bibfnamefont{V.~V.} \bibnamefont{Yashchuk}},
  \bibnamefont{and}
  \bibinfo{author}{\bibfnamefont{M.}~\bibnamefont{Zolotorev}},
  \bibinfo{journal}{Phys. Rev. A} \textbf{\bibinfo{volume}{62}},
  \bibinfo{pages}{043403} (\bibinfo{year}{2000}).

\bibitem[{\citenamefont{Novikova and Welch}(2002)}]{NovikovaW02}
\bibinfo{author}{\bibfnamefont{I.}~\bibnamefont{Novikova}} \bibnamefont{and}
  \bibinfo{author}{\bibfnamefont{G.~R.} \bibnamefont{Welch}},
  \bibinfo{journal}{J. Mod. Opt.} \textbf{\bibinfo{volume}{49}},
  \bibinfo{pages}{349} (\bibinfo{year}{2002}).

\bibitem[{\citenamefont{Vanier et~al.}(1998)\citenamefont{Vanier, Godone, and
  Levi}}]{vanier98}
\bibinfo{author}{\bibfnamefont{J.}~\bibnamefont{Vanier}},
  \bibinfo{author}{\bibfnamefont{A.}~\bibnamefont{Godone}}, \bibnamefont{and}
  \bibinfo{author}{\bibfnamefont{F.}~\bibnamefont{Levi}},
  \bibinfo{journal}{Phys. Rev. A} \textbf{\bibinfo{volume}{58}},
  \bibinfo{pages}{2345} (\bibinfo{year}{1998}).

\bibitem[{\citenamefont{Kitching et~al.}(2002)\citenamefont{Kitching, Knappe,
  and Hollberg}}]{hollberg'02}
\bibinfo{author}{\bibfnamefont{J.}~\bibnamefont{Kitching}},
  \bibinfo{author}{\bibfnamefont{S.}~\bibnamefont{Knappe}}, \bibnamefont{and}
  \bibinfo{author}{\bibfnamefont{L.}~\bibnamefont{Hollberg}},
  \bibinfo{journal}{Appl. Phys. Lett.} \textbf{\bibinfo{volume}{81}},
  \bibinfo{pages}{553} (\bibinfo{year}{2002}).

\bibitem[{\citenamefont{Merimaa et~al.}(2003)\citenamefont{Merimaa, Lindvall,
  Tittonen, and Ikonen}}]{merimaa'03}
\bibinfo{author}{\bibfnamefont{M.}~\bibnamefont{Merimaa}},
  \bibinfo{author}{\bibfnamefont{T.}~\bibnamefont{Lindvall}},
  \bibinfo{author}{\bibfnamefont{I.}~\bibnamefont{Tittonen}}, \bibnamefont{and}
  \bibinfo{author}{\bibfnamefont{E.}~\bibnamefont{Ikonen}},
  \bibinfo{journal}{J. Opt. Soc. Am. B} \textbf{\bibinfo{volume}{20}},
  \bibinfo{pages}{273} (\bibinfo{year}{2003}).

\bibitem[{\citenamefont{Vanier et~al.}(2003)\citenamefont{Vanier, Levine,
  Janssen, and Delaney}}]{Vanier03}
\bibinfo{author}{\bibfnamefont{J.}~\bibnamefont{Vanier}},
  \bibinfo{author}{\bibfnamefont{M.~W.} \bibnamefont{Levine}},
  \bibinfo{author}{\bibfnamefont{D.}~\bibnamefont{Janssen}}, \bibnamefont{and}
  \bibinfo{author}{\bibfnamefont{M.~J.} \bibnamefont{Delaney}},
  \bibinfo{journal}{IEEE Trans.\ Instrum.\ Meas.}
  \textbf{\bibinfo{volume}{52}}, \bibinfo{pages}{822} (\bibinfo{year}{2003}).

\bibitem[{\citenamefont{Novikova et~al.}(2001)\citenamefont{Novikova, Matsko,
  and Welch}}]{novikova'01ol}
\bibinfo{author}{\bibfnamefont{I.}~\bibnamefont{Novikova}},
  \bibinfo{author}{\bibfnamefont{A.~B.} \bibnamefont{Matsko}},
  \bibnamefont{and} \bibinfo{author}{\bibfnamefont{G.~R.} \bibnamefont{Welch}},
  \bibinfo{journal}{Opt. Lett.} \textbf{\bibinfo{volume}{26}},
  \bibinfo{pages}{1016} (\bibinfo{year}{2001}).

\bibitem[{\citenamefont{Mair et~al.}(2002)\citenamefont{Mair, Hager, Phillips,
  Walsworth, and Lukin}}]{phillips03}
\bibinfo{author}{\bibfnamefont{A.}~\bibnamefont{Mair}},
  \bibinfo{author}{\bibfnamefont{J.}~\bibnamefont{Hager}},
  \bibinfo{author}{\bibfnamefont{D.~F.} \bibnamefont{Phillips}},
  \bibinfo{author}{\bibfnamefont{R.~L.} \bibnamefont{Walsworth}},
  \bibnamefont{and} \bibinfo{author}{\bibfnamefont{M.~D.} \bibnamefont{Lukin}},
  \bibinfo{journal}{Phys. Rev. A} \textbf{\bibinfo{volume}{65}},
  \bibinfo{pages}{031802} (\bibinfo{year}{2002}).

\end{thebibliography}
\end{document}